\documentstyle[aps,11pt,preprint]{revtex}

\newcommand{\bfq}{\bf q}
\newcommand{\pbar}{\bar{p}} 

\newcommand{\af}{\alpha} 
\newcommand{\sbar}{\bar{s}} 
\newcommand{\tbar}{\bar{t}} 
\newcommand{\twoba}{\bar{2}} 
\newcommand{\thrba}{\bar{3}} 
\newcommand{\lam}{\lambda} 
\newcommand{\Ht}{\lambda_{\bar{2}}\lambda_{4};\lambda_{1}\lambda_{\bar{3}}} 
 
\newcommand{\Jx}{J_{\xi}} 
\newcommand{\lamf}{\lam_{\bar{2}}\lam_{4}} 
\newcommand{\lami}{\lam_{1}\lam_{\bar{3}}} 
\newcommand{\Mx}{M_{\xi}} 

\begin{document}
\title{The $\xi$(2230) Meson and The Pomeron Trajectory }   
\author{L.\ C.\ Liu}
\address{Theoretical Division, T-2, Los Alamos National Laboratory \\
  Los Alamos, NM 87545, USA}
\author{Wei-hsing Ma} 
\address{Institute of High Energy Physics, Academia Sinica \\
 Beijing, China} 
\date{October 23, 1999} 
\maketitle
\begin{abstract}
We examine the possibility that the $\xi(2230)$ meson is a member of   
the Pomeron trajectory. A method of connecting      
the $\xi\rightarrow p\pbar$ decay width and the $pp$ cross sections
through the Pomeron residue function is presented. We have used a   
relativistic, singularity-free form factor to make the analytic   
continuation of the residue function between crossed channels.   
We predict that if the $\xi(2230)$ meson is a Pomeron, then 
it should have a $\xi\rightarrow p\pbar$ decay width of about   
2 MeV.

\end{abstract}

\vspace{1in} 


\pagebreak  
\narrowtext
Prior to the advent of the quantum chromodynamic (QCD) theory, the 
Regge theory was extensively used in describing the strong interaction.  
In spite of its success, the nature of the Pomeron, which plays a
crucial role in explaining the asymptotic behavior of the hadron-hadron 
interaction, remains unclear.  
Over the years, precise phenomenological fits to
high-energy data have determined that\cite{Blo}\cite{Pdb1}   
the spin and the mass of the Pomerons  
satisfy the relation    
$ \af(t)= \af(0) + \af' t$\  
with $\af(0)\simeq 1.08$
and $Re[\af']\simeq 
0.2 $(GeV)$^{-2}$. If the Pomeron($\cal P$)
 trajectory is more than a theoretical device,
then it should have member constituencies identifiable with physical states.  
To date, no such states have been found.  

In this report we show that the    
$\xi(2230)$ meson could be a good candidate for Pomeron.  
This meson was first observed in the radiative decay of the $J/\Psi$ to  
$K\bar{K}$ by the MARK III collaboration\cite{Mark}.  
Interest in $\xi(2230)$ resurged\cite{Mey} when the 
BES collaboration\cite{Bai} measured the reaction $J/\Psi\rightarrow 
\gamma\xi$ and determined   
that this state has a mass of 2230 MeV with a total width $\Gamma_{tot}= 20\pm 17$ MeV.  
Its spin is still uncertain\cite{Euro}, being either 
$2^{++}$ or $4^{++}$, but will be determined in future measurements. 

If $J=2$ is confirmed, then   
with $J=Re[\af(t)]=2$ at   
$t=M_{\xi}^{2}=(2.23)^2=5 $(GeV)$^{2}$ 
the $\xi$ satisfies the above-mentioned spin-mass 
relation of the Pomeron.  
In the following we shall go beyond       
this relation and show what would be the constraint on  
the $\xi\rightarrow p\pbar$ decay width, $\Gamma_{\xi\rightarrow p\pbar}$,  
when $\xi$ is on the ${\cal P}$-trajectory.
In particular, we show how we can relate it   
to the residue function of the Pomeron. 
We then use this relation and the measured $pp$  
cross sections to predict the width. 
In our analysis, we have used for the first time     
a relativistic and singularity-free form factor.  
We emphasize that our analysis  
does not rely on any assumption about the subhadronic content  
of the $\xi$ meson. Hence, it is model-independent of the latter.  
The signifcance of our result will be elucidated.  

Let us examine the 
$pp$ elastic scattering (denoted $12\rightarrow 34$) 
mediated by the exchange of 
the $\xi(2230)$ meson. 
The corresponding $t$-channel process  
is, therefore, the $p\pbar$ scattering (denoted 
$1\bar{3}\rightarrow \bar{2} 4)$ via the formation and decay of the $\xi$. 
Let $m(0.94$ GeV),  
$\Mx(2.23$ GeV), and $\Jx$ be, respectively, the mass of the $p(\pbar)$,   
the mass of the $\xi$, and the spin of the $\xi$.  
The $t$-channel helicity Feynman amplitude 
can be written as  
\begin{equation} 
 M_{\Ht}(\sbar,\tbar)= 4m^{2} C_{I}  
[4\pi(2\Jx+1)]<\lamf\mid{\cal M}^{\Jx}(\sbar)\mid \lami> 
d_{\lam\lam'}^{\Jx}(z_{t})   
\label{eq:1} 
\end{equation}  
where $ \lam\equiv \lam_{1}-\lam_{\thrba}, 
\lam'\equiv \lam_{\twoba}-\lam_{4}$, and $C_{I}=1/2$ is the isospin 
factor. For clarity, we denote      	 
the square of the total c.m. energy and the momentum transfer in 
the $s$-channel by $s$ and $t$, respectively, and those in the $t$-channel 
by $\sbar=(p_{1}+p_{\bar{3}})^{2}$ and 
$\tbar=(p_{1}-p_{\bar{2}})^{2}$. In the c.m. of the $p\pbar$, 
 $\sbar=4(m^{2}+k_{t}^{2}), 
\tbar=-2k_{t}^{2}(1-cos\theta_{t})\equiv -2k_{t}^{2}(1-z_{t})$.  
Because $p_{\bar{3}}=-p_{3}, -p_{\bar{2}}=p_{2}$, we have $\sbar=t$ 
and $\tbar=s$. In Eq.(\ref{eq:1})  
\begin{equation} 
<\lamf\mid {\cal M}^{\Jx}(\sbar)\mid \lami> 
=\frac{ G_{\lam'}H_{\lamf;\Jx}(\sbar)\ G_{\lam}H_{\Jx;\lami}(\sbar)}
{\sbar - M^{2}_{\xi} + iM_{\xi}\Gamma_{tot} } ,  
\label{eq:2} 
\end{equation} 
The $H$ denotes
the form factor for the $\xi p\pbar$ vertex in the helicity basis  
and $G$ the coupling constant.  
  
The Regge-pole amplitude due to the ${\cal P}$-trajectory can be written as\cite{Pdb1} 
\begin{equation} 
 A_{\Ht}(\sbar,\tbar)= C_{I}  
\frac{-4\pi^{2} (2\af +1)\beta_{\lam\lam'}(t) 
(-1)^{\af+\lam}\frac{1}{2}[1+(-1)^{\af}]}{sin\pi(\af+\lam')}   
d_{\lam\lam'}^{\af}(z_{t})\equiv  
{\cal A}^{t}_{\lam\lam'}\ ,   
\label{eq:7} 
\end{equation} 
where $\beta_{\lam\lam'}(t)$ 
stands for $\beta_{\lamf;\lami}(\sbar\equiv t)$. 
The contribution of the $\xi$ meson to the ${\cal P}$-trajectory is obtained by 
calculating their overlap as follows:  
\begin{equation} 
\frac{1}{2}\int_{-1}^{+1} M_{\Ht}(\sbar,\tbar)d_{\lam\lam'}^{\Jx}(z_{t})dz_{t}
=lim_{\af\rightarrow\Jx} \frac{1}{2}
 \int_{-1}^{+1} A_{\Ht}(\sbar,\tbar)
d_{\lam\lam'}^{\Jx}(z_{t})dz_{t}\ . 
\label{eq:8} 
\end{equation} 
By using  
$\int_{-1}^{+1} d^{\Jx}_{\lam\lam'}(z_{t})d^{\Jx}_{\lam\lam'}(z_{t})dz_{t}
= 2/(2\Jx+1)$  and the following relation, which we have derived,   
\begin{equation}   
\frac{1}{2}\int_{-1}^{+1} d_{\lam\lam'}^{\af}(z_{t})d_{\lam\lam'}^{\Jx}(z_{t})dz_{t}
= \frac{sin\pi\af}{\pi(\af-\Jx)(\af+\Jx+1)} {\cal F}
\label{eq:9} 
\end{equation} 
with 
\begin{equation} 
{\cal F}\equiv (-1)^{\Jx}\left(  
\frac{(\af-\lam)!(\Jx+\lam)!}{(\af+\lam)!(\Jx-\lam)!}\right)^{\frac{1}{2}} ,  
(\lam=0, \pm1),  
\label{eq:10} 
\end{equation}
we obtain          
\begin{eqnarray} 
 & & \frac{16m^{2}\pi G^{2}H_{\lamf;\Jx}(t)H_{\Jx;\lami}(t)} 
{t - \Mx^{2} + i\Mx\Gamma_{tot} }  \nonumber \\    
& = & \left[ \frac{-4\pi(2\af+1)\beta_{\lam,
\lam'}(t)}{(\af-\Jx)(\af+\Jx+1)}
\frac{(-1)^{\af+\lam}\frac{1}{2}[1+(-1)^{\af}]sin\pi\af}  
{sin\pi(\af+\lam')}\ {\cal F}\right]_{\af\rightarrow \Jx} \nonumber \\  
& =&  \frac{-4\pi \beta_{\lam\lam'}(t_{r})/\af'_{R}}{t - t_{r} + i \af_{I}(t_{r})/\af'_{R}}\ . 
\label{eq:11} 
\end{eqnarray}  
Hence, for even $\Jx$ we have the identifications $t_{r}=\Mx^{2},\ 
\af_{I}(t_{r})/\af_{R}'= \Mx\Gamma_{tot}$, and 
\begin{equation} 
\beta_{\lam\lam'}(t_{r})= -\af_{R}' 4m^{2} G_{\lam'}H_{\lamf;\Jx}(t_{r})
G_{\lam}H_{\Jx;\lami}(t_{r})\ . 
\label{eq:12} 
\end{equation} 
In obtaining Eq.(\ref{eq:11}) we have  
defined $\af_{R}\equiv Re(\af), 
\af_{I}\equiv Im(\af)$, and have used the Taylor expansion 
$\af(t) 
= \af_{R}(t_{r}) + (\af_{R}' +i\af_{I}'(t_{r}))(t-t_{r}) + \af_{I}(t_{r})  
+ ...$, with $\af_{R}(t_{r})\equiv \Jx$, $\af_{I}\ll \af_{R},$ 
and $\af_{I}'\ll 
\af_{R}'.$  

The diagonal matrix element of Eq.(\ref{eq:12}) is directly related  to 
the $\xi\rightarrow p\pbar$ decay width. Using the method of   
\cite{Hai}, we obtain
\begin{eqnarray} 
\frac{1}{2}\Gamma_{\xi\rightarrow p\pbar} &=& 
C_{I}\frac{m^{2}}{M^{2}_{\xi}}\frac{\mid {\bf q_{r}}\mid}{16\pi^{2}} 
\sum_{\lam_{p}\lam_{\pbar}}\mid G_{\lam}
H_{\lam_{p}\lam_{\pbar}}({\bf q_{r}}^{2})\mid^{2} \nonumber \\ 
 &=& C_{I} \frac{\mid{\bf q_{r}}\mid}{64\pi^{2}\alpha'_{R}M^{2}_{\xi}} 
\sum_{\lam} \beta_{\lam\lam}(t_{r})\ , 
\label{eq:3} 
\end{eqnarray}     
where   
${\bf q_{r}}^{2}=t_{r}/4 -m^{2}=0.36$ (GeV)$^{2}$ 
and $\lam=\lam_{p}-\lam_{\pbar}$. The ${\bfq_{r}}$ is the value of the   
c.m. momentum ${\bfq}$ of the $p\pbar$ system 
at $\sbar=M^{2}_{\xi}\equiv t_{r}$.    
On the other hand, the $\beta_{\lam\lam'}(t)$ at $t\leq 0$ is related to the
$pp$ cross sections. 
The spin-averaged total $pp$ cross section is 
given by 
\begin{equation} 
\sigma^{pp}_{tot}  =    
\left(\frac{1}{4}\right)\left(\frac{4\pi}{k_{s}}\right)  
\frac{1}{8\pi k_{s}}\sum_{\lam_{3}\lam_{4};\lam_{1}\lam_{2}}Im[{\cal A}^{s}
_{\lam_{3}\lam_{4};\lam_{1}\lam_{2}}(s,t)]_{\mid_{t=0}}\ ,   
\label{eq:13a} 
\end{equation}
where $k_{s}$ and ${\cal A}^{s}$ are the $s$-channel c.m. momentum
and amplitudes, and $k^{2}_{s}\simeq s/4$. Furthermore,     
$\sum_{\lam_{3}\lam_{4};\lam_{1}\lam_{2}}Im[{\cal A}^{s}
_{\lam_{3}\lam_{4};\lam_{1}\lam_{2}}] =    
 \sum_{\Ht}Im[{\cal A}^{t}_{\Ht}]$    
due to crossing symmetry. There are sixteen  
helicity amplitudes for $p\pbar$ scattering.     
However, only five amplitudes are independent as a result of   
the parity, total spin conservation, and time-reversal invariance.
We denote them as ${\cal A}^{t}_{i}$.  
The indices $i=1,...,5$ correspond successively to      
$(\Ht)=$ $(++;++),$ $(++;--),$ $(+-;+-),$ $(+-;-+),$ $(++;+-)$ 
with $\pm$ denoting, respectively, the helicities $\pm\frac{1}{2}$. 
In the limit $s\rightarrow\infty$ and, hence, $z_{t}\rightarrow\infty$
\cite{Pdb1}
\begin{equation} 
d^{\af}_{\lam\lam'} \sim \frac{(-1)^{(\lam-\lam')/2}(2\af)!(z_{t}/2)^{\rho}} 
{\sqrt{(\af+M)!(\af-M+\mid\lam-\lam'\mid)!(\af-M)!(\af+M-\mid\lam-\lam'\mid)!}}
\label{eq:14} 
\end{equation} 
and 
\begin{equation} 
\left(\frac{z_{t}}{2}\right)^{\rho} \sim (-1)^{\rho}
\left(\frac{s}{4m^{2}-t}\right)^{\rho}  
\label{eq:15} 
\end{equation} 
where $\rho\equiv\af-M+\mid\lam-\lam'\mid/2+\mid\lam+\lam'\mid/2$
and $M\equiv max(\mid \lam\mid,\mid \lam'\mid)$.  
Upon introducing Eqs.(\ref{eq:14}) and (\ref{eq:15}) into Eq.(\ref{eq:7}), 
one sees that  
${\cal A}^{t}_{1}={\cal A}^{t}_{2}$ and  ${\cal A}^{t}_{3}=-{\cal A}^{t}_{4}$.  
Consequently,    
\begin{equation} 
\sigma^{pp}_{tot} = \frac{1}{2s}Im[4{\cal A}^{t}_{1} + 
8{\cal A}^{t}_{5}]\mid_{t=0}\ .    
\label{eq:16} 
\end{equation}
The $pp$ elastic differential cross section is given by 
\begin{equation} 
\frac{d\sigma^{pp}}{dt}=\frac{1}{16\pi s^{2}}
\left( 4\mid {\cal A}^{t}_{1}\mid^{2}   
 + 4\mid {\cal A}^{t}_{3}\mid^{2}   
 + 8\mid {\cal A}^{t}_{5}\mid^{2} \right) \ .  
\label{eq:16b}     
\end{equation} 
In the above equations, ${\cal A}^{t}_{1}, {\cal A}^{t}_{3}, {\cal A}^{t}_{5}$ 
depend, respectively, on $\beta_{00}(t), \beta_{11}(t), \beta_{10}(t)$.    

The $\beta(t)$ can be calculated from $\beta(t_{r})$ with the use of a model
that we specify below. First, we note that in deriving Eq.(\ref{eq:12}) 
one only requires $\lim_{t\rightarrow t_{r}}\af_{R}(t)=\Jx $. 
From the mathematical point of view,  the position of $t$ at 
which the limit is taken can be anywhere along 
the Regge/Pomeron trajectory. Hence, we have       
the functional equality  
\begin{equation} 
\beta_{\lam\lam'}(t)= -\af_{R}' 4m^{2} G_{\lam'}H_{\lamf;\Jx}(t)
G_{\lam}H_{\Jx;\lami}(t)  
\label{eq:12b} 
\end{equation}
along this trajectory. In fact,  
the theory of analytic functions implies that the equality exists  
in a region where both $\beta$ and $H$ are analytic functions.
In general the functional form of $H(t)$ 
can also depend on $t$.  However, because   
there are no bound states and other resonances 
at $t <  M^{2}_{\xi}$,   
we can assume that the functional form 
of $H(t)$ is the same in the entire     
region $t \leq M^{2}_{\xi}$.  
From Eqs.(\ref{eq:12}) and (\ref{eq:12b}) we have      
\begin{equation} 
\beta_{\lam\lam'}(t)= \beta_{\lam\lam'}(t_{r})
\frac{H_{-\lam_{4},\lam_{4};\Jx}(t)
H_{\Jx;\lam_{1},-\lam_{1}}(t)} 
{H_{-\lam_{4},\lam_{4};\Jx}(t_{r})
H_{\Jx;\lam_{1},-\lam_{1}}(t_{r})}\ .     
\label{eq:17} 
\end{equation}

Once the form factor $H$ is known, 
the $\beta_{\lam\lam'}$ can be calculated
from their values at $t_{r}$ which, by Eq.(\ref{eq:3}),
are related to the decay width.    
It is advantageous to model 
the form factor in the $LS-$basis because in this latter basis the form    
factor has a well-known $q^{L}$ threshold behavior that can be  
explicitly incorporated into the model. 
Because the helicity basis is related to the $LS-$basis by a unitary 
transformation\cite{Wick}, one has   
$\sum_{\lam} \mid G_{\lam} H_{\lam_{p},\lam_{\pbar};\Jx}
\mid^{2} =  
\sum_{LS} \mid g_{LS}F_{LS}\mid^{2}$ in Eq.(\ref{eq:3}).  
Here      
$F_{LS}$ and $g_{LS}$ denote, respectively, the form factor and 
the dimensionless coupling constant in the $LS-$basis.  
For $p\pbar$ system with $J^{P}=2^{+}$, the parity conservation leads to  
$L=1, 3$ and $S=1$ only. We will henceforth    
omit the subscript $S$.

In the literature, the most commonly used hadronic form factor is either   
of an exponential form or of a multipole form.   
However, these form factors are unsuitable to 
analyses involving channel-crossing.  
For example, the exponential form factor $exp(t/\Lambda^{2})$ is analytic in 
the $s$-channel where $t\leq 0$ but diverges in the $t$-channel where 
$t>0$. The multipole form factor   
$[\Lambda^{2}/(\Lambda^{2}-t)]^{n}\  (n=1,2,...)$ has no pole on the real 
axis of $t$  
in the $s$-channel (where $t\leq 0$) but will have it   
in the $t$-channel (where $t>0$). Conversely, if we use  
$exp(-t/\Lambda^{2})$ or  
$[\Lambda^{2}/(\Lambda^{2}+t)]^{n}$, then the situation will be reversed.  
We propose the following   
form factor which is singularity free: 
\begin{equation} 
\mid F_{L}(t)\mid^{2} =\left( \frac{t/4-m^{2}}{{\bf q_{r}}^{2}} \right)^{L}   
\left(\frac{e^{t_{r}/\lam_{t}^{2}}}{R(-x_{t})+e^{t/\lam_{t}^{2} }}\right)^{2} 
\left( 
\frac{1+e^{-t_{r}/\lam_{s}^{2}} }{R(x_{s})+e^{-t/ \lam_{s}^{2} }}\right)^{2}
\label{eq:18} 
\end{equation} 
where $(t/4-m^{2})^{L}\equiv f_{L} ={\bf q}^{2L}$ reflects      
the $q^{L}$-dependence of the $F_{L}$.   
Because $F_{L}(t_{r})\equiv 1$ we have from Eq.(\ref{eq:3}) and from  
the relation between $H$ and $F$ that $\Gamma_{\xi\rightarrow p\pbar}\propto
 g_{1}^{2}+g_{3}^{2}$.  

In Eq.(\ref{eq:18}),   
$x_{s}\equiv (t-2m^{2})/\lam_{s}^{2}$ and $x_{t}\equiv 
(t-2m^{2})/\lam_{t}^{2}$. 
The function $R$ is analytic and is defined by
\begin{equation} 
R(x)=\frac{1}{2}(1+tanh(ax))=\frac{e^{ax}}{e^{ax}+e^{-ax}}\ .  
\label{eq:19} 
\end{equation} 
It rapidly changes from 0 to 1 when $x $ changes from $<0$ to 
$>0$, with $a$ controling   
the transition speed at $x=0$. With $a>10$, $R$ will be very
close to a step function but does not have the discontinuity of the latter. 
Consequently, the form factor is a continuous function of $t$. 
In the $s$-channel ($t\leq 0$),
$\mid F(t)\mid^{2} \propto    
f_{L}\  [1+exp(t/\lam_{t}^{2})]^{-2}[exp(-t/\lam_{s}^{2})]^{-2}  
\sim t^{L}\ exp(2t/\lam_{s}^{2})$, which goes to 0 as $t\rightarrow -\infty$.  
Hence, $\lam_{s}$ controls the form factor. Since $F(t)$ does not have the 
physical-channel energy $s$ as an explicit variable, it does not diverge with 
$s$. When $t$ reaches the $t$-channel physical domian     
($t > 4m^{2}$), $\mid F(t)\mid^{2} \propto   
 f_{L}\ [exp(t/\lam_{t}^{2})]^{-2}[1+exp(-t/\lam_{s}^{2})]^{-2}
\sim t^{L}\  exp(-2t/\lam_{t}^{2}) \rightarrow 0 $ when $t\rightarrow \infty$, 
exhibiting the correct energy behavior in the $t$-channel. Here in the    
$t$-channel the $\lam_{t}$ controls the form factor.  
The above well-behaved $t$-dependence of $F(t)$ makes the latter a good tool 
for continuing $\beta$ between the direct and crossed channels.  

Eqs.(\ref{eq:3}) and (\ref{eq:16}) to (\ref{eq:17}) 
allow us to predict the decay width $\Gamma_{\xi\rightarrow p\pbar}$ 
from the measured $pp$ total and elastic cross sections, and vice versa. 
We have determined the form factor parameters from the experimental   
$pp$ total and elastic cross sections data\cite{Blo}\cite{Euro} 
at $\sqrt{s}=53$ and 62 GeV. As anticipated,  
$\lam_{s}$ is mainly determined by the diffraction peak of the $d\sigma/dt$.  

The result of our fit is given in Table  \ref{tab:1} where the wider parameter 
ranges at 62 GeV are due to the larger experimental error bars. 
Table  \ref{tab:1} indicates that if $\xi$ 
lies on the ${\cal P}$-trajectory, then it has a $\Gamma_{\xi\rightarrow  
p\pbar}$ between 1.5 and 2 MeV. This decay width,  
in addition to spin, can be used to check whether $\xi$ is a Pomeron. 
We  note that the BES collaboration\cite{Bai} cited nearly equal 
branching ratios for the four observed decays modes. Using these   
equal branching ratios and the above $\Gamma_{\xi\rightarrow p\pbar}$ 
we conclude that the $\Gamma_{tot}$ for the $\xi$ is at least 8 MeV.   
While this lower bound is compatible with the published data, it is, however, 
necessary to ascertain in future experiments that no other  
important decay channels than those observed in ref.\cite{Bai} are left out.  

We recall that the theoretical modeling of the Pomeron started in the  
perturbative QCD sector\cite{Low}--\cite{Bal}. Various gluon-exchange models 
were proposed. However, all these models gave an $\alpha(0)$ much greater than 
the phenomenologcial value of 1.08. 
Recently, there is a growing interest in  
a possible connection between the Pomeron and the glueball\cite{Kis}. 
We emphasize that at this time there is no 
convincing proof that Pomeron is a glueball. As to the   
$\xi(2230)$ meson itself, it       
can either be a tensor glueball   
or a $q\bar{q}-$glue mixture\cite{Clo}--\cite{Pag}. 
Frank Close\cite{Clo3} has shown that 
if $\xi$ is a tensor glueball then its total width $\Gamma_{tot}$ 
would be of the order of 25 MeV; but no discussion was made on the 
$\Gamma_{\xi\rightarrow p\pbar}$.  
Although the $\Gamma_{tot}$ of the BES result\cite{Bai} falls within this limit,
that measurement needs to be improved\cite{Mey}.  
We believe that high-statistics   
data on many decay modes of all the tensor states   
in this mass region are needed to pin down the gluonic content of the $\xi$.   
In this work, we do not study the microscopic composition of the $\xi$. 
Instead, we investigate the conditions that $\xi$ could be a Pomeron.    
If the Pomeron 
status of the $\xi$ is established, then the 
subhadronic structure of the $\xi$   
will be directly relevant to that of the Pomeron.   
In this respect, our study will help clarify  
the Pomeron-glue connection.

In summary, we have derived the relation 
between the residue function of the    
Pomeron trajectory and decay widths of its member states.   
Meaurements of the spin and the 
$p\pbar$-decay width of the $\xi$ are important for  
determining if the $\xi(2230)$ meson could be the long-sought Pomeron. 
 
We thank Drs. J.C. Peng,  L.S. Kisslinger, P.R. Page, and L. Burakovsky       
for helpful discussions.   

\begin{table}
\caption{The $\Gamma_{\xi\rightarrow p\pbar} $ predicted from the measured $pp$ 
cross sections}  
\label{tab:1}
\begin{tabular}{|c|c|c|c|c|c|}
$\sqrt{s}$ (GeV) & $\lam_{s}$ (GeV)\ \ \ 
 & $\lam_{t}$ (GeV)\ \ \ & $g_{1}$ \ \ \ & $g_{3}$\ \ \  &
$\Gamma_{\xi\rightarrow p\bar{p}}$ (MeV)\ \ \  \\  
\tableline
 53   &  0.65--0.67 & 3.43--3.44 & 1.13--1.14 & 0.295--0.299  & 1.85--1.86\\ 
\tableline 
 62   &  0.66--0.68 & 3.25--4.07 & 1.07--1.19 & 0.277--0.360  & 1.65--2.08\\ 
\end{tabular}
\end{table}
\pagebreak

\end{document}